\documentclass[journal]{IEEEtran}
\usepackage{amsopn,amssymb,amsfonts}
\usepackage[cmex10]{amsmath}
\usepackage{subfigure}
\usepackage{graphicx}
\usepackage{algorithm}
\usepackage{algorithmic}
\usepackage{epstopdf}
\usepackage[normalem]{ulem}
\usepackage{bm}

\newcommand{\figref}[1]{Fig.\,\ref{#1}}

\newcounter{mytempeqcounter}

\title{Performance Analysis of Molecular Spatial Modulation (MSM) in Diffusion based Molecular MIMO Communication Systems}

\author{Tayyebeh~Jahani-Nezhad,
	and Foroogh~S.~Tabataba
\thanks{Tayyebeh~Jahani-Nezhad is with the Department
	of Electrical and Computer Engineering, Isfahan University of Technology, Isfahan,
	84156-83111, Iran e-mail: (t.jahaninezhad@ec.iut.ac.ir).}
\thanks{Foroogh~S.~Tabataba is with the Department
	of Electrical and Computer Engineering, Isfahan University of Technology, Isfahan,
	84156-83111, Iran e-mail: (fstabataba@cc.iut.ac.ir).}}

\begin{document}
\maketitle

\begin{abstract}
In diffusion-based molecular communication, information is transferred from a transmitter to a receiver using molecular carriers. The low achievable data rate is the main disadvantage of diffusion-based molecular over radio-based communication. One solution to overcome this disadvantage is molecular MIMO communication. In this paper, we introduce molecular spatial modulation (MSM) in molecular MIMO communication to increase the data rate of the system. Also, special detection methods are used, all of which are based on the threshold level detection method. They use diversity techniques  in molecular communication systems if the channel matrix that we introduce is full rank. Also, for a 2$\times$1 system, we define an optimization problem to obtain the suitable number of molecules for transmitting to reduce BER of this systems. Then the proposed modulation is generalized to $2\times2$ and $4\times4$ systems. In each of these systems, special detection methods based on the threshold level detection are used.  Finally, based on BER, systems using MSM are fairly compared to the systems that have similar data rates. The simulation results show that the proposed modulation and detection methods reduce BER. Whereas the proposed methods are very simple and practical for molecular systems.

\vspace{0.8 em}\emph{\textbf{Index Terms:}}  Molecular Communication Systems, Diffusion, Molecular Spatial Modulation,  Threshold Level Detection,  Molecular MIMO Communication, Data Rate, Convex Optimization.
\end{abstract}

\section{Introduction}
\IEEEPARstart{N}{owadays} nanotechnologies are developing and affecting many areas. The presence of nanotechnology topics in communications can expand the boundaries of this science. In specific medical and industrial applications, nanoscale communications are needed. As a result, a new and interdisciplinary field is presented called molecular communication \cite{RN109}, \cite{RN113}.
In the diffusion-based molecular communication model, information is coded into the concentration, type,
or release time of the molecules. The molecules released into the environment by a transmitter, travel from the transmitter to a receiver based on the Brownian motion mechanism \cite{RN111}, \cite{RN112}. The low achievable data rate is the main disadvantage of diffusion-based molecular over radio-based communication and there are not some researches on this \cite{MIMO_prototype}.  One solution to overcome this disadvantage is to use molecular MIMO communication \cite{32}.
In the most of the researches done on the molecular communication systems, only one transmitter and one receiver are used and rarely researches can be seen based on the multi-transmitter or multi-receiver antennas. Molecular MIMO communication systems have been first introduced in \cite{32} in which different diversity techniques have been discussed in the presence of multi-user interference and by ignoring inter symbol interference (ISI). In \cite{32}, there is no discussion about the dependence of the paths between the transmitters and the receivers. The authors only assumed that each receiver antenna is independent of the others if they are far enough apart.
In \cite{MIMO_prototype}, the molecular communication system with two transmitters and two receivers has been proposed. Then, four detection methods have been presented based on the channel matrix which has been introduced in this work. Also, the authors have mentioned that  the first-hitting probability equation in single antenna scheme cannot be used in multi-antenna molecular communication systems  due  to  the  dependency of  the different  paths. \cite{effect} has investigated the effects of interferences for different modulation techniques in a system with two receivers and transmitters. \\ In \cite{2recv}, the effect of adding a new receiver on the absorption probability, the channel capacity of the communication link and  bit error probability has been studied. \cite{monte} has used the Monte-Carlo simulation to plot the absorption probability of molecules in terms of time and distance in a molecular communication system with one transmitter and multi receivers which are located in the same distance from the transmitter. In \cite{coop}, a cooperative molecular communication system has been presented. In this work, one transmitter and multi receivers have been considered. Each receiver makes a local hard decision about the transmitted bit in the current time slot and then sends this decision to a fusion center using a special type of molecules. Then,  using a fusion rule, the fusion center merges the local
hard decisions to get a global decision. The extended version of this paper has been presented in \cite{optimize} in which the error performance of the mentioned system is analyzed and optimized.
In \cite{learning}, the author introduces a machine learning method for modeling the molecular MIMO channel in a $2\times 2$ MIMO system.
\cite{spatial} studies spatial diversity techniques in diffusion-based molecular MIMO communication system. In this work, the Alamouti and repetition MIMO coding are suggested and analyzed in a symmetric $2\times 2$ MIMO system in which the channel coefficients are achieved by a trained artificial neural network.
In \cite{synch},  a blind clock synchronization mechanism in a SIMO system using only one symbol is proposed. In this synchronization the channel coefficients are unknown.  To estimate the channel impulse response of the molecular MIMO systems, the maximum likelihood and the least-squares estimators are studied in \cite{CIRestim}, and to minimize the
Cramér-Rao bound, the training sequences are designed.
None  of  above  studies  have  considered  spatial  modulation in their works. The most related reference to our work is \cite{SMMC} in which molecular spatial  modulation has been studied in a parallel and independent work.  In this study,  the spatial modulation has been characterized for a molecular MIMO system with a different topology. In this work, the SISO channel model is used while due  the dependency of the different paths, this channel model may not be proper for molecular communication with multiple receivers \cite{MIMO_prototype}, \cite{monte}, \cite{spatial}, \cite{CIRestim}. In \cite{SMMC}, the equal gain combining (EGC) is used to detect the concentration symbol. Because of the high complexity of this detector, the EGC may be impractical for nano-scale devices like the maximum likelihood detector \cite{Communication over Diffusion-Based}.
\\ In this paper, we first introduce a proper matrix called channel matrix for molecular MIMO systems. Then, we examine the rank of this matrix in different scenarios. We would use diversity techniques in molecular MIMO communication if the channel matrix is full rank. In the next step, we introduce molecular spatial modulation (MSM) to increase the data rate of the system.
Then, in a $2\times1$ system based on MSM to reduce bit error rate, a convex optimization problem is proposed in which the optimum number of transmitting molecules according to system structure is calculated. In the next step, we generalize the proposed modulation for $2\times2$ and $4\times4$ systems. In
each of these systems, special detection methods are used, all of which are based on the threshold level detection method
and use diversity techniques in molecular communication systems. Finally, based on BER, systems using MSM are fairly compared with the systems that have similar data rates. According to the simulation results, the proposed modulation and detection methods increase the data rate and reduce BER. In addition, the proposed methods are very simple.
\\The rest of this paper is organized as follows: In Section II, the background of molecular communication is expressed. Some properties of molecular MIMO communication systems is introduced in section III, we detail the proposed molecular spatial modulation in section IV. Section V presents the simulation results and the last section represents the conclusion.
\section{Molecular Communication Background}
When a molecule is released from a transmitter, it propagates randomly in the fluid to arrive at the receiver. The arrival time of the molecule is random due to its free motion. This random propagation time is called the first arrival time. If the fluid does not have drift velocity, the distribution of the first arrival time will be in the form of a L\'{e}vy distribution and will be as follows \cite{comprehen}
\begin{align}
\begin{aligned}
{f_T}(t) = \left\{ {\begin{array}{*{20}{c}}
	{ \frac{{{R_r}}}{{d + {R_r}}}\sqrt {\frac{\lambda }{{2\pi {t^3}}}} \exp ( - \frac{\lambda }{{2t}})},&\hspace{0.5cm} t > 0\\\\
	0 ,&\hspace{0.5cm}t \le 0
	\end{array}} \right.
\label{eqchegal}
\end{aligned}
\end{align}
where $R_r$, $d$, and $t$ are the receiver radius, the transmitter-receiver distance and the first arrival time, respectively. $\lambda$ is also determined as $\lambda  = \frac{{{d^2}}}{{2D}}$ where $D$ is the diffusion coefficient.
As a result, integrating ${f_T}(t)$, the probability of hitting of molecules until the desired time $T$ is obtained as
\begin{align}
\begin{aligned}
{p_k}= \int\limits_{0}^{{T}} {{f_T}(\tau )d\tau}= \frac{{{R_r}}}{{{R_r} + d}}\textnormal{erfc}(\frac{d}{{\sqrt {4D{T} } }}),
\label{eqProb}
\end{aligned}
\end{align}
where $\textnormal{erfc}(x)$ is the complementary error function \cite{3dim}.  As a consequence, the number of absorbed molecules by the receiver in $k$th time slots can be approximated by a Binomial random variable as
\begin{align}
\begin{aligned}
N_k^{Rx} \sim \textnormal{Binomial}(N_k^{Tx},p_k),
\end{aligned}
\end{align}
where $N_k^{Rx}$, $N_k^{Tx}$ denotes the number of received molecules at receiver ($Rx$) and the number of emitted molecules from the transmitter ($Tx$). If the number of trials is large, the Normal distribution can be used as an approximation of the Binomial distribution with same mean and variance \cite{Papoulis2002}. In the molecular communication, the number of emitted molecules from the transmitter is large enough that we can use Normal distribution instead of Binomial \cite{energy}, \cite{RN114}. It is because that the using the Normal distribution is easier than using Binomial distribution. As a result, the number of captured molecules by the receiver is
\begin{align}
\begin{aligned}
N_k^{Rx}\sim{\mathcal {N}}(N_k^{Tx}{p_k},N_k^{Tx}{p_k}(1 - {p_k})),
\label{NormalVar}
\end{aligned}
\end{align}
where ${\mathcal {N}}(\mu,\sigma^2)$ denotes a normal random variable with $\mu$ as mean and $\sigma^2$ as variance.
\begin{algorithm}
	\caption{pseudocode for computing
		absorption probabilities at spherical receiver} \label{alg:RANSAC}
	
	\begin{algorithmic}[1]
		\REQUIRE $M$: Number of molecules to be emitted\\
		$\Delta t$: Simulation step time\\
		$R_r$: Radius of a spherical receiver\\
		$T_s$: Symbol interval\\
		$D$: Diffusion coefficient\\
		$[x_t,y_t,z_t]$: Emission point (Transmitter node coordinates)\\
		$[x_r,y_r,z_r]$: Receiver node coordinates\\
		$[x,y,t]$: Molecule location coordinates\\
		$N_{R_x}$: Number of captured molecules by receiver\\
		$d_L$: The limit of distance\\
		\ENSURE Molecule absorption probability $p$.
		\STATE Compute simulation end time $(T_E)$ using $\Delta t$ and $T_s$
		\FOR {$m=1$ to $M$}
		\FOR{$t = 1$ to $T_E$}
		\IF {time for emission }
		\STATE $ [x,y,t]\leftarrow[x_t,y_t,z_t]$
		\ELSE
		\STATE Compute $\Delta x,\Delta y,\Delta z \sim\;{\cal N}(0,2D\Delta t)$
		\STATE  Molecules propagation \cite{MoMotion}: \\$x \leftarrow x+ \Delta x$ \\
		$y \leftarrow y+ \Delta y$\\$z \leftarrow z+ \Delta z$
		\ENDIF
		\IF {$\left\| {(x,y,z) - ({x_r},{y_r},{z_r})} \right\| < {R_r}$}
		\STATE The molecule is absorbed and $N_{R_x}\leftarrow N_{R_x}+1$
		\ELSIF {$\left\| {(x,y,z) - ({x_r},{y_r},{z_r})} \right\| > d_L$}
		\STATE break;
		\ELSE
		\STATE Return to step 4
		\ENDIF
		\ENDFOR
		\ENDFOR
		\STATE  $p\leftarrow N_{R_x}/M $
	\end{algorithmic}
	
\end{algorithm}
\section{ Molecular MIMO systems}\label{III}
\begin{figure}
	\centering
	\includegraphics[width=0.8\columnwidth]{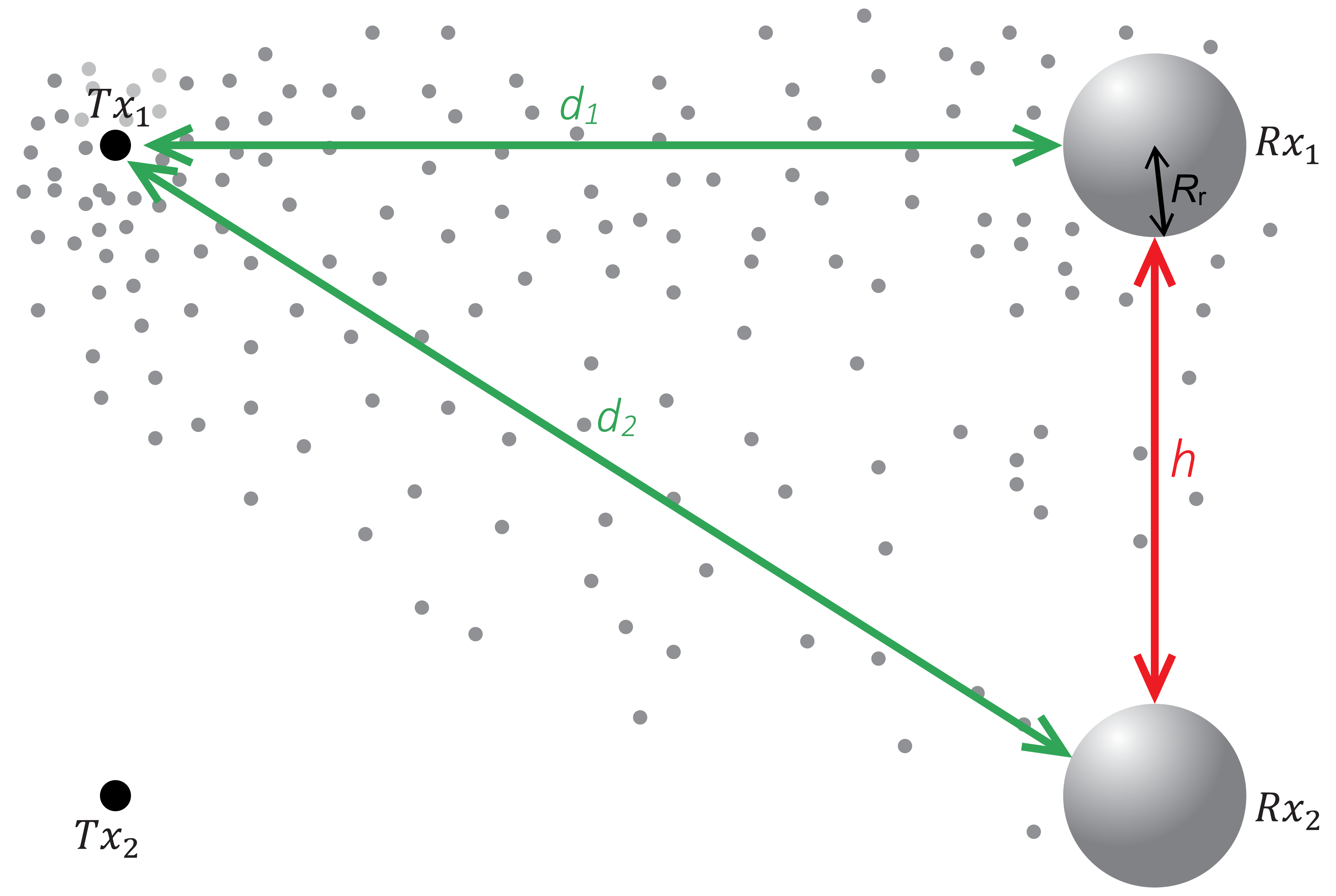}
	\caption{A molecular communication system with two transmitters and two spherical receivers.}
	\label{2Tx2Rx}
	\centering
\end{figure}

\begin{figure}

	\centering
	\includegraphics[width=1\columnwidth]{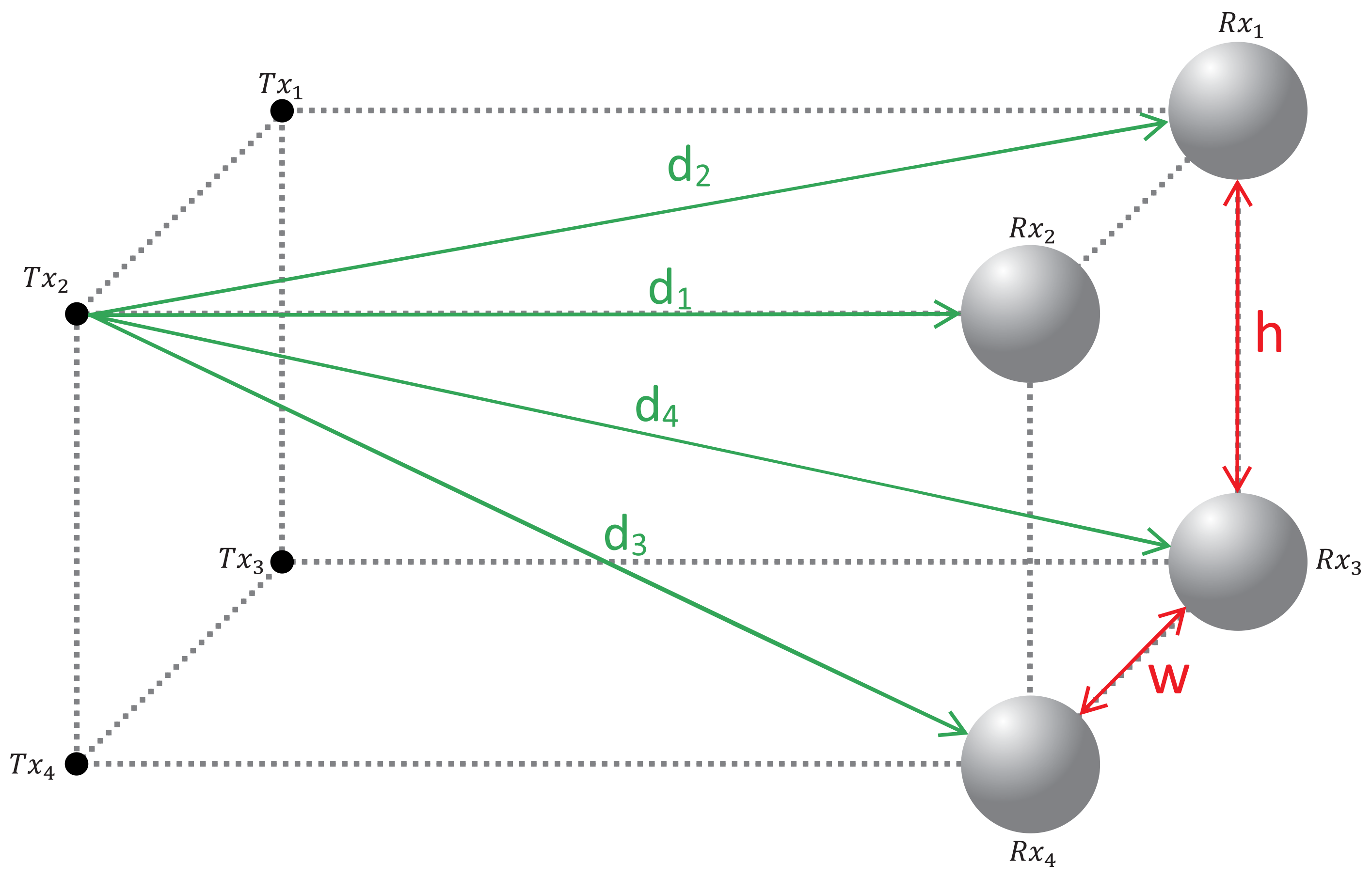}
	\caption{A molecular communication system with four transmitters and four spherical receivers.}
	\label{asd}

\end{figure}

\begin{figure}
	\centering
	\includegraphics[width=1\columnwidth]{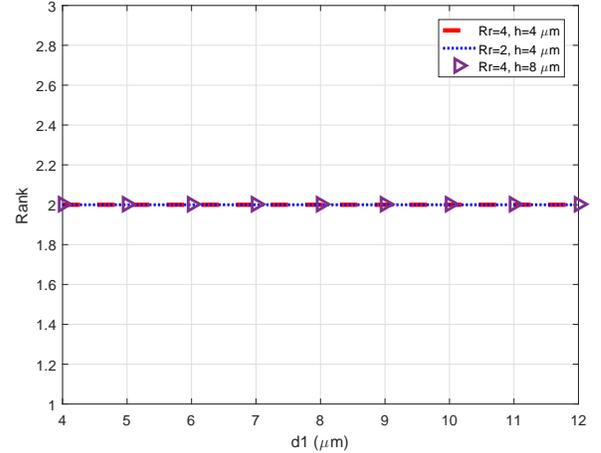}
	\caption{The channel matrix rank of 2$\times$2 molecular system in terms of different distances.}
	\label{Rank2}
	\centering
\end{figure}

\begin{figure}
	\centering
	\includegraphics[width=1\columnwidth]{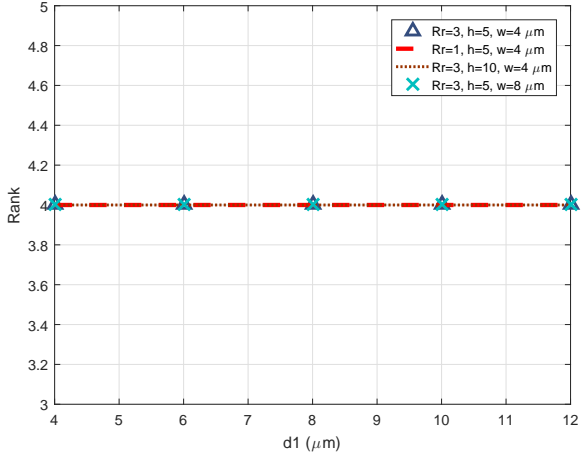}
	\caption{The channel matrix rank of 4$\times$4 molecular system in terms of different distances.}
	\label{Rank4}
	\centering
\end{figure}

As discussed in the previous section, one way to increase the data rate of molecular communication systems is the use of multiple transmitters and multiple receivers. The probability of hitting of molecules in the 1$\times$1 molecular system in the $k$th time slot is
\begin{align}
\begin{aligned}
{p_k}^{Tx_j} = \frac{{{R_r}}}{{{R_r} + {d_j}}}(\text{erfc}(\frac{{d_j}}{{\sqrt {4Dk{T_s} } }})-\text{erfc}(\frac{{d_j}}{{\sqrt {4D(k-1){T_s} } }})),
\label{eqProbTs}
\end{aligned}
\end{align}
where $d_j$ is the distance between $Tx_j$ and the receiver. Also, $T_s$ is the length of each time slot.
When another receiver is added to the 1$\times$1 molecular system, this equation is not proper for hitting probability of molecules in 1$\times$2 or other molecular communication with multiple receivers \cite{MIMO_prototype}. In \cite{MIMO_prototype} a model function similar to molecular SISO
in a three-dimensional environment \eqref{eqProbTs} with some control coefficients is proposed  for hitting probability. The model function is as follows
\begin{equation}
{p^{(i,j)}}(t) = \frac{{{b_1}{R_r}}}{{{d_{ij}} + {R_r}}}\text{erfc}(\frac{{{d_{ij}}}}{{{{(4D)}^{{b_2}}}{t^{{b_3}}}}}),
\label{controli}
\end{equation}
where  ${{d_{ij}}}$ is the distance between $j$th transmitter and $i$th receiver. $b_1$, $b_2$, and $b_3$ are control coefficients which are obtained from the simulation of Brownian motion and fitting the result of hitting probability to \eqref{controli}. Two molecular MIMO systems are proposed in \figref{2Tx2Rx} and \figref{asd}. For hitting probability of molecules of these systems the equation \eqref{controli} can be used.
\begin{figure*}[!t]
		\normalsize
	\includegraphics[width=2\columnwidth]{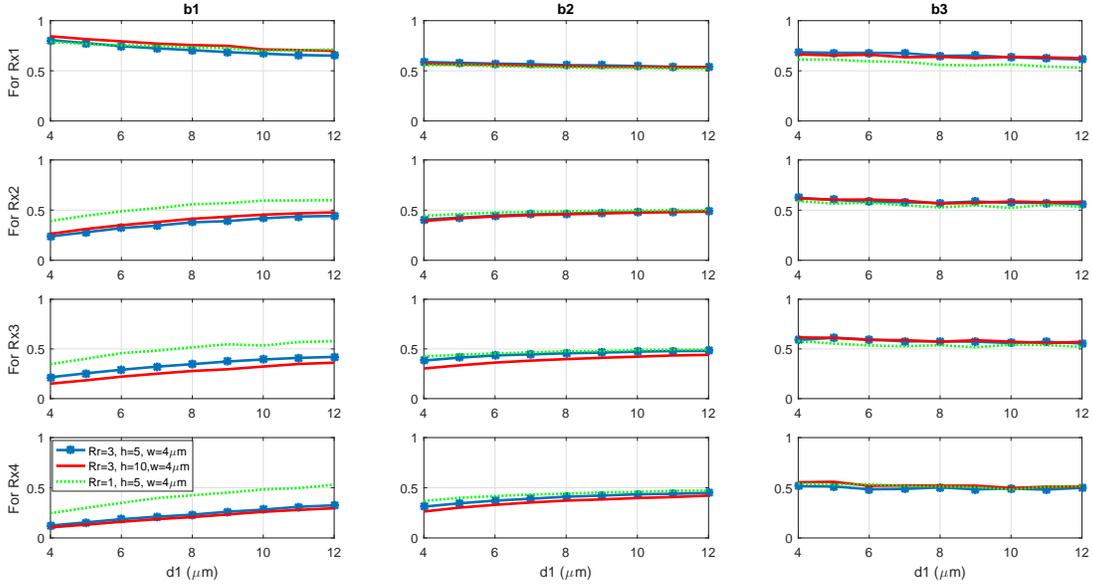}
	\caption{Control coefficients ${b_1}$, ${b_2}$, and ${b_3}$ in terms of $d_1$ for different $R_r$ and $h$ for 4$\times$4 MIMO scheme. }
	\label{bi_plotss}
	\centering

	\vspace*{4pt}
\end{figure*}
Thus, the control coefficients should be calculated. For the 4$\times$4 system, like \figref{asd}, the control coefficients are obtained using simulation as described in algorithm \ref{alg:RANSAC}.\\
The differences between the fitted model and the simulated values actually are computed with root-mean-square error (RMSE) measure and the order of error is about $10^{-4}$. \figref{bi_plotss} shows the control coefficients ${b_1}$, ${b_2}$ and ${b_3}$ in terms of $d_1$ for different $R_r$, $h$ and $w$ for 4$\times$4 MIMO system, where $h$ and $w$ are the distance between two  receivers in vertical and horizontal direction, respectively.  \\
The existence of multiple receivers in a system means the existence of different propagation paths. Sending same data across the different propagation path is called diversity and causes more reliability. We will use diversity techniques in molecular MIMO communication systems if the channel matrix (which is defined in \eqref{cc4}) is full rank.  By defining the number of transmitted molecules from each transmitter as input and the number of captured molecules in each receiver as output, the channel matrix can be introduced. For example in 4$\times$4 system, the channel matrix is given in \eqref{cc4}.
\begin{align}
\begin{aligned}
&\textbf{y} = \left[ {\begin{array}{*{20}{c}}
	{{N^{R{x_1}}}}\\
	{{N^{R{x_2}}}}\\
	{{N^{R{x_3}}}}\\
	{{N^{R{x_4}}}}
	\end{array}} \right] =\\
&\left[ {\begin{array}{*{20}{c}}
	{{\delta ^{(1,1,{x_1})}}}&{{\delta ^{(1,2,{x_2})}}}&{{\delta ^{(1,3,{x_3})}}}&{{\delta ^{(1,4,{x_4})}}}\\
	{{\delta ^{(2,1,{x_1})}}}&{{\delta ^{(2,2,{x_2})}}}&{{\delta ^{(2,3,{x_3})}}}&{{\delta ^{(2,4,{x_4})}}}\\
	{{\delta ^{(3,1,{x_1})}}}&{{\delta ^{(3,2,{x_2})}}}&{{\delta ^{(3,3,{x_3})}}}&{{\delta ^{(3,4,{x_4})}}}\\
	{{\delta ^{(4,1,{x_1})}}}&{{\delta ^{(4,2,{x_2})}}}&{{\delta ^{(4,3,{x_3})}}}&{{\delta ^{(4,4,{x_4})}}}
	\end{array}} \right]
\left[ {\begin{array}{*{20}{c}}
	{x_1}\\
	{x_2}\\
	{x_3}\\
	{x_4}
	\end{array}} \right]= \textbf{H}\textbf{x},
\label{cc4}
\end{aligned}
\end{align}
where ${{N^{R{x_i}}}}$ and ${x_j}$ are the number of captured molecules by the $i$th receiver and the released molecules from the $j$th transmitter. Each element of the channel matrix is a random variable as follows
\begin{equation}
\delta^{(i,j,{x_j})} \sim {\frac{1}{{{x_j}}}}\textnormal{Binomial}({x_j},p^{(i,j)}).
\label{bbb}
\end{equation}
Where $p^{(i,j)}$ is the probability of capturing molecule by the $i$th receiver  which is transmitted from the $j$th transmitter in the current time slot and is obtained from \eqref{controli}. Each element of the channel matrix given in \eqref{bbb} includes the randomness of the molecule's diffusion.\\
Simulation can also be used to create this matrix in the 4$\times$4 system. In other words in each level of simulation and for a structure with certain dimensions, several molecules are released according to the sent bit 0 or 1 in the transmitter. These molecules propagate randomly to be placed in the vicinity of the receiver and absorbed by the receiver. Accordingly, using several simulations and sending molecules from all of the transmitters,  a channel matrix is obtained, in which each entry shows that how possible it is for a molecule released by the $j$th transmitter to be absorbed by the  $i$th receiver. Also, it should be considered that to create this matrix, it is assumed that each of the receivers is able to detect the transmitter after receiving molecule. \\In other words, it is assumed that each transmitter uses a special kind of molecule.
For different distances $d$, $h$, $w$, and $R_r$,  the channel matrix is obtained and the rank of this matrix can be computed. If the channel matrix is full rank, the independence of the paths between the transmitters and the receivers is concluded. In \figref{Rank2}  and  \figref{Rank4} the rank of channel matrices in the proposed  2$\times$2 and 4$\times$4  molecular MIMO communication system are shown respectively.
From the above simulations, the channel matrices in these two systems are always full rank. Notice that  the number of released molecules from the transmitters are not zero in these two systems. Otherwise, if all transmitters send bit 0, it is clear that the channel matrix will not be full rank.  This property will be used in the next sections.
\section{System Model and Analysis of MSM schemes}
\begin{figure*}[!t]
		\normalsize
	\centering
	\includegraphics[width=2\columnwidth]{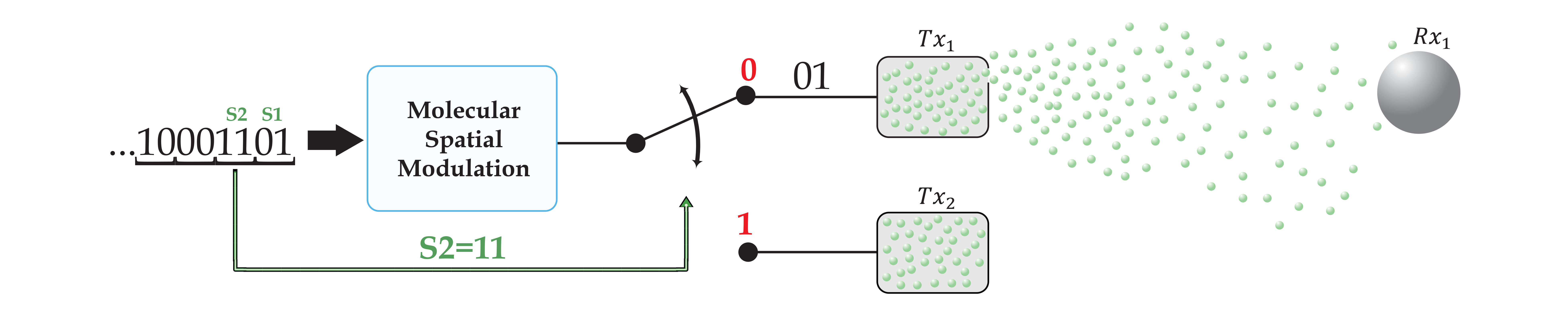}
	\caption{View of a molecular communication system based on MSM.}
	\label{MSM}
	\centering
	\vspace{4pt}
\end{figure*}

\IEEEpubidadjcol
Spatial modulation is used in  the systems that have multiple receivers and multiple transmitters \cite{SpatialRef}. The system is designed to be time-slotted with intervals of length $T_s$. The transmitter releases its molecules at the beginning of each time slot.   Firstly, in this section, a system with two transmitters and one receiver is discussed, then the system is expanded in the next section. The transmitters are modeled as  point sources of molecules and the receiver is modeled as a spherical absorbing one. In each time slot, one of the transmitters sends one bit of information to the receiver. Each transmitter uses binary concentration shift keying (BCSK) modulation to transmit bit 0 or bit 1 \cite{effect}. We consider $L_0$ and $L_1$ as the number of molecules released to send bit~0 and bit~1, respectively.\\
In each time slot, only one transmitter sends its information. Depending on which transmitter is sending, a bit is detected in the receiver.  Therefore, the bit stream is divided into blocks each containing two bits. The first bit determines which transmitter is active to transmit a bit. This means choosing a point in the spatial constellation and in this topology, the constellation has two spatial points. The second bit represents the bit of information that  must be sent by the active transmitter. This modulation is called \textit{Molecular Spatial Modulation} or MSM. This method avoids inter-link interference (ILI) but there is the inter-symbol interference (ISI). In this paper, it is assumed that only the last symbol has a significant ISI effect over the current symbol like \cite{effect}, \cite{energy}, \cite{Arj}. Note that in this modulation, both of $L_0$ and $L_1$ should be non-zero values; otherwise, the channel matrix is not full rank.  In \figref{MSM} the view of MSM based molecular communication system is shown.
\begin{figure}
	\centering
	\includegraphics[scale=.06]{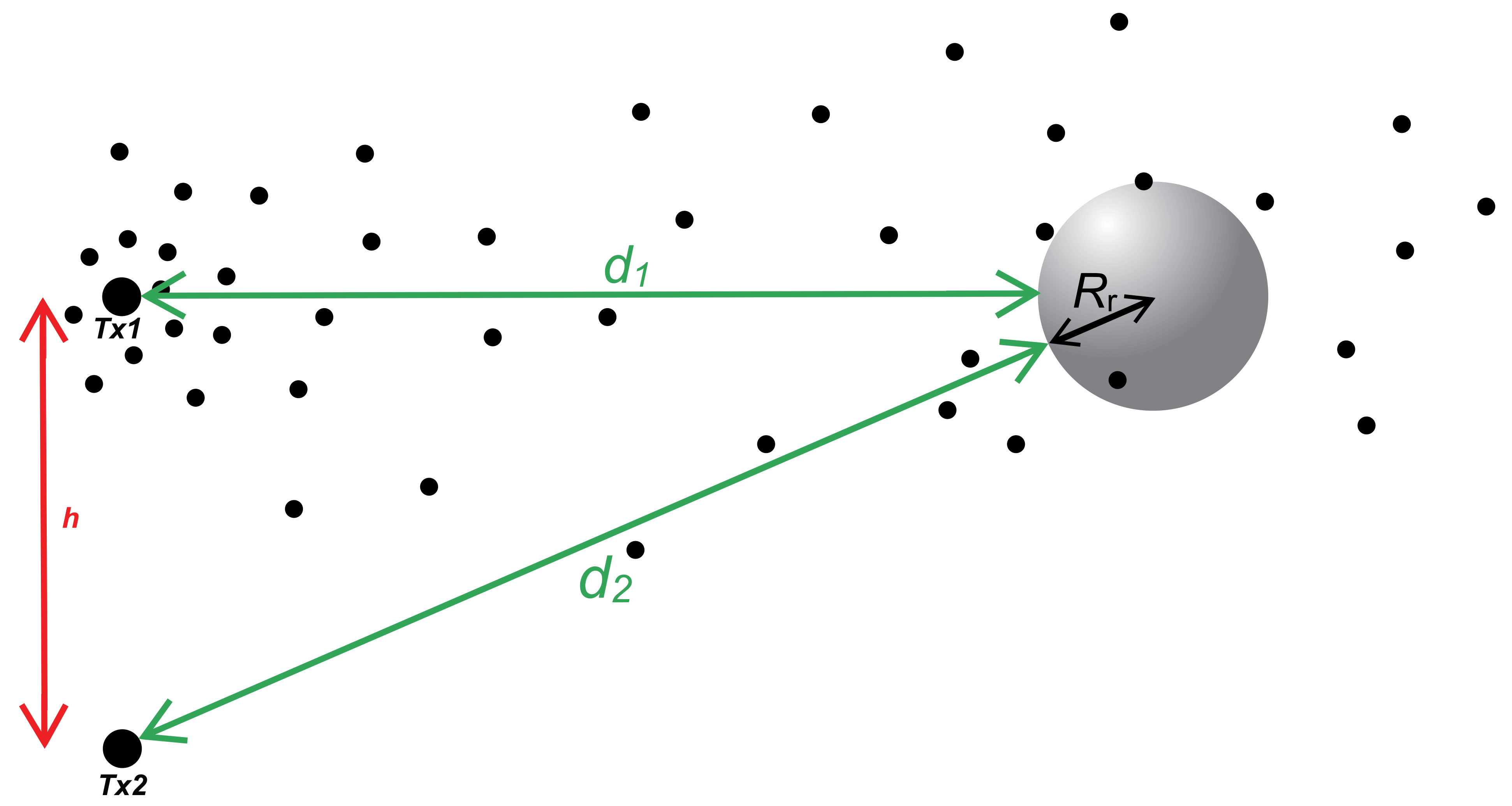}
	\caption{A molecular communication system with two transmitters and a spherical receiver.}
	\label{2Tx}
	\centering
\end{figure}
\subsection{Analysis of 2$\times$1 MSM system}
\subsubsection{Proposed modulation scheme}
In  this  section,  we consider  a  2$\times$1  MSM system  as  shown  in \figref{MSM}. The two transmitter
antennas are placed $h$ distance apart. The probability of molecule hitting receiver which is released from the $j$th transmitter ($Tx_j$) in $k$th time slots is obtained from \eqref{eqProbTs}.
The number of molecules absorbed by the receiver among all of the molecules released from each transmitter obeys \eqref{NormalVar}. Considering the ISI caused by the previous symbol, the total number of molecules captured by the receiver has the following distribution
\begin{align}
\begin{aligned}
&{N_k}^{Rx}\sim  \mathcal{N}(\mu ,{\sigma ^2}),\\
&\mu  = {N_k}^{T{x_j}}{p_1}^{T{x_j}} + {N_{k - 1}}^{T{x_i}}{p_2}^{T{x_i}},\\
&{\sigma ^2} = {N_k}^{T{x_j}}{p_1}^{T{x_j}}(1 - {p_1}^{T{x_j}}) + {N_{k - 1}}^{T{x_i}}{p_2}^{T{x_i}}(1 - {p_2}^{T{x_i}}),
\label{Sig&mean}
\end{aligned}
\end{align}
where ${p_2}^{T{x_i}}$ and ${p_1}^{T{x_j}}$ are given in \eqref{eqProbTs}. Note that  if the number of trials is large, Normal distribution can be used instead of Binomial with the same mean and variance \cite{Papoulis2002}. In molecular communication, the number of emitted molecules from the transmitter is large enough to use Normal distribution instead of Binomial. It is because that using the Normal distribution is easier than using Binomial distribution. Also, in \eqref{Sig&mean}, the special characteristic of the normal random variable, i.e., the sum of two independent normal random variables is normal, is used \cite{Papoulis2002}. Also, ${N_k}^{T{x_j}}$ and ${N_{k-1}}^{T{x_i}}$ can get two values of $L_0$ or $L_1$ based on the selected bit to be sent in each time slot.
In the following, to reduce bit error rate (BER), a convex optimization problem is designed in which the optimum number of transmitting molecules according to the system structure is calculated. We denote the convex optimization problem as
\begin{align}
\begin{aligned}
&\mathop {{\mathop{\rm minimize}\nolimits} }\limits_{{L_0},{L_1}}
\hspace{0.2cm} ||\bm{\mu} - \textbf{b}||_2^2 +||{\bm{\sigma ^2}} - \textbf{c}||_2^2 \\
&subject\hspace{0.1cm}to\hspace{0.8cm}{L_0} + {L_1} = {L_{total}},
\label{optimization}
\end{aligned}
\end{align}
where $L_{total}$ is the sum of total molecules which are released to send bit 0 and bit 1. $\bm{\mu}$ and ${\bm{\sigma ^2}}$ are the vectors which contain all the calculated means and variances for different sending modes in two consecutive time slots. In other words $\bm{\mu}$  and ${\bm{\sigma ^2}}$ are defined as follows
\begin{align}
\begin{array}{l}
{\mu\hspace{-1mm}=\hspace{-1mm}{[{\mu _{(00,00)}}, {\mu _{(01,00)}}, {\mu _{(10,00)}}, {\mu _{(11,00)}},..., {\mu _{(10,11)}}, {\mu _{(11,11)}}]^T},}\\\\
{\sigma ^2}\hspace{-1mm}=\hspace{-1mm}{[\sigma _{(00,00)}^2, \sigma _{(01,00)}^2, \sigma _{(10,00)}^2, \sigma _{(11,00)}^2,..., \sigma _{(10,11)}^2, \sigma _{(11,11)}^2]^T},\\
\end{array}
\end{align}
where $S_p$ and $S_c$ in ${\mu _{(S_p,S_c)}}$ or ${\sigma _{(S_p,S_c)}^2}$ denote previous and current symbols which are transmitted. ${(.)^T}$ is an operator which computes the transpose of vectors. Also, $\textbf{b}$ and $\textbf{c}$ are the desired vectors. For example, the definition of \textbf{b} is
\begin{equation}
\textbf{b} = {[{b_{00}},{b_{00}},{b_{00}},{b_{00}},{b_{01}},...,{b_{11}},{b_{11}}]^T}.
\label{bbbb}
\end{equation}
The optimization problem goal is that regardless of the previous symbol, the number of received molecules for the current symbol is approximately equal. Thus, all four consecutive elements of \textbf{b} are chosen equal to each other. In \eqref{bbbb}, $b_{ij}$ denotes the desired average  number of molecules which are received in the receiver when the symbol $ij$ is transmitted. Vector \textbf{c}  has the same definition as vector \textbf{b}. By solving the optimization problem, the suitable number of molecules
for sending bit 0 and bit 1 are obtained. Following that only four different values for the mean and variance of Normal distributions are calculated using \eqref{Sig&mean}.
\subsubsection{Detection}
\begin{figure}
	\centering
	\includegraphics[width=1\columnwidth]{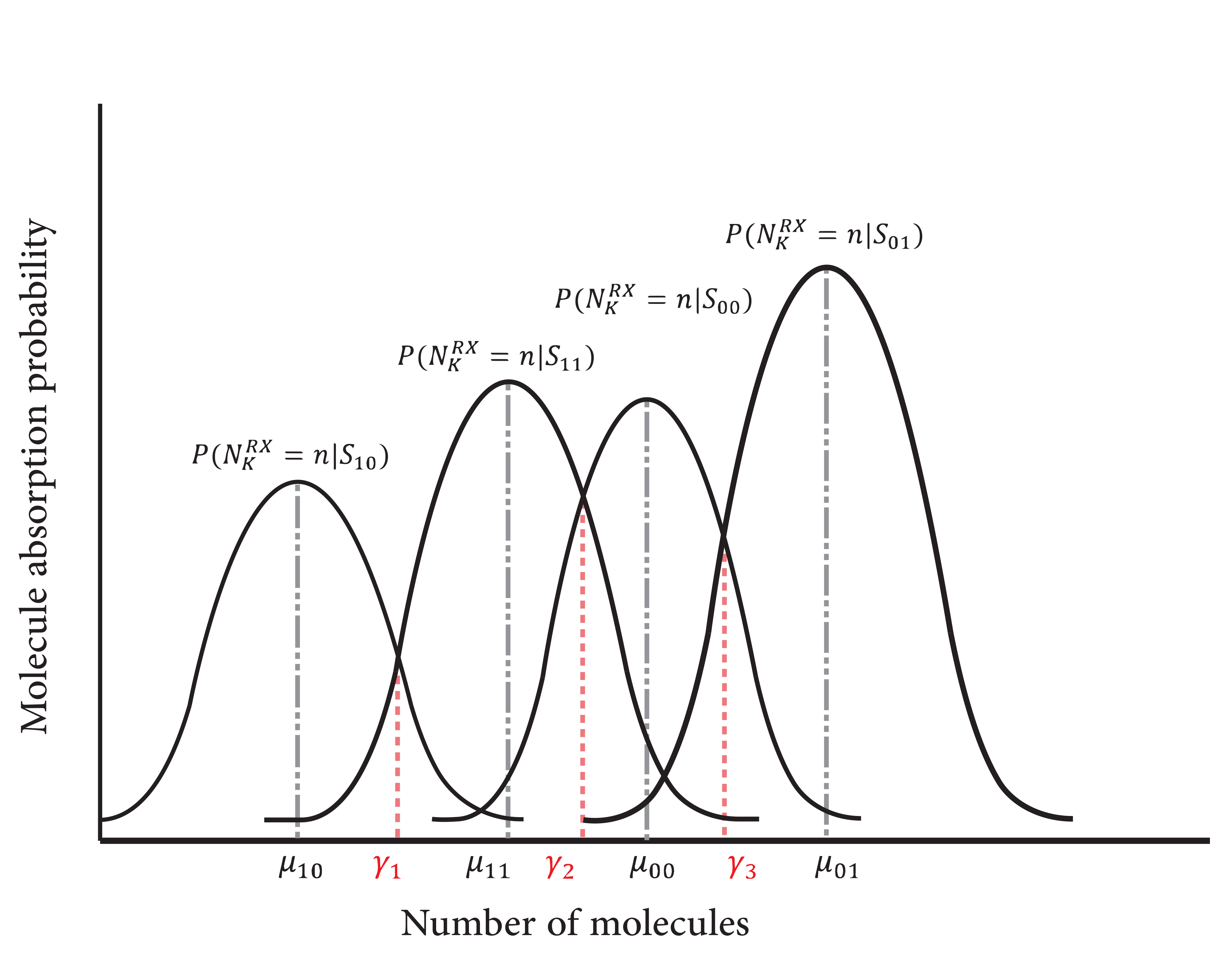}
	\caption{An example of finding threshold levels when MSM is used in 2$\times$1 or 2$\times$2 molecular systems.}
	\label{Dist}
	\centering
\end{figure}

For symbol detection at the receiver side, the threshold levels are needed. These threshold levels are obtained from the intersection points of two distributions \cite{signal}. This leads to the equality
\begin{align}
\begin{aligned}
\frac{1}{{\sqrt {2\pi {\sigma ^2}_{(S_p,S_c)}} }}&\exp ( - \frac{{{{(\gamma  - {\mu _{(S_p,S_c)}})}^2}}}{{2{\sigma ^2}_{(S_p,S_c)}}}) =\\
&\frac{1}{{\sqrt {2\pi {\sigma ^2}_{(S'_p,S'_c)}} }}\exp ( - \frac{{{{(\gamma  - {\mu _{(S'_p,S'_c)}})}^2}}}{{2{\sigma ^2}_{(S'_p,S'_c)}}}),
\end{aligned}
\end{align}
where $\gamma$ is the threshold level.
Finally, this equality becomes in the form of a quadratic equation as follows
\begin{align}
\begin{aligned}
A{\gamma ^2} + B\gamma  + C = 0,
\end{aligned}
\end{align}
where
\begin{align}
\begin{aligned}
A &= {\sigma ^2}_{(S_p,S_c)} - {\sigma ^2}_{(S'_p,S'_c)},\\
B &= 2({\mu _{(S_p,S_c)}}{\sigma ^2}_{(S'_p,S'_c)} - {\mu _{(S'_p,S'_c)}}{\sigma ^2}_{(S_p,S_c)}),\\
C &= ({\mu _{(S'_p,S'_c)}}{\sigma ^2}_{(S_p,S_c)} - {\mu _{(S_p,S_c)}}{\sigma ^2}_{(S'_p,S'_c)})\\
&- ({\sigma ^2}_{(S'_p,S'_c)}{\sigma ^2}_{(S_p,S_c)})\ln(\frac{{{\sigma ^2}_{(S_p,S_c)}}}{{{\sigma ^2}_{(S'_p,S'_c)}}}),
\end{aligned}
\end{align}
The threshold value is the positive root of the quadratic equation above.
As a result, four threshold levels  like \figref{Dist} are calculated.
The receiver compares the total number of received molecules in each time slot with the threshold levels and then makes a decision on the current symbol.
\subsection{Analysis of  2$\times$2 MSM system}\label{2x2Sec}
\subsubsection{Proposed modulation scheme}
To increase achievable data rate and reduce bit error rate, the proposed modulation is generalized to the  2$\times$2 system.
Consider \figref{2Tx2Rx} in which, there are two transmitters and two receivers. According to \ref{III}, the molecule hitting probability in the systems with more than one receiver is calculated using \eqref{controli} and
in any system with special structure, $b_i$ values are obtained using repetitive simulations.
In the mentioned system, same as before, only one transmitter sends one bit of information in each time slot. Therefore, inter-link interference or ILI will not occur. Therefore, considering which transmitter sent the information, a bit in the receiver will be detected. In addition, each transmitter can send either bit 0 or bit 1. This causes detecting another bit in the receiver. In other words, two bits are sent in each time slot. Also, we assume that the ISI is caused by the previous symbol, as we did before. The number of molecules captured by each receiver is
\begin{align}
\begin{aligned}
&{N_1^{R{x_1}}} = \delta_1^{(1,1,{x_1})}{x_1}{a_1} + \delta_1^{(1,2,{x_2})}{x_2}{a_2} + {I_1},\\
&{N_1^{R{x_2}}} = \delta_1^{(2,1,{x_1})}{x_1}{a_1} + \delta_1^{(2,2,{x_2})}{x_2}{a_2} + {I_2},
\label{NumberOfRec}
\end{aligned}
\end{align}
In these equations, $I_i$ denotes the ISI effect and $x_i$ is the number of molecules which are sent by the $i$th transmitter. $x_i$ can get either $L_0$ or $L_1$. Also, $a_j$ is the $j$th transmitter activation parameter. If the transmitter is active, this parameter gets 1 and if not, it gets 0. $\delta_1^{(i,j,{x_j})}$ is a random variable as follows
\begin{align}
\begin{aligned}
\delta _1^{(i,j,{x_j})} \sim \frac{1}{{x_j}}\textnormal{Binomial}({x_j},p_1^{(i,j)}),
\label{SigmaBino}
\end{aligned}
\end{align}
which is normalized Binomial random variable with $x_j$ trials and success probability $p_k^{(i,j)}$. This random variable shows the  molecule absorption probability which is released by the $j$th transmitter in the $k$th time slot and is absorbed by the $i$th receiver. $p_k^{(i,j)}$ can also be calculated as follows
\begin{align}
\begin{aligned}
{p_k^{(i,j)}}& = \frac{{{b_1}{R_r}}}{{{d_{ij}} + {R_r}}}(\text{erfc}(\frac{{{d_{ij}}}}{{{{(4D)}^{{b_2}}}{({kT_s})^{{b_3}}}}})\\
&-\text{erfc}(\frac{{{d_{ij}}}}{{{{(4D)}^{{b_2}}}{{((k-1)T_s)}^{{b_3}}}}})).
\end{aligned}
\end{align}
According to \eqref{NumberOfRec}, the relationship between the input and the output of this MIMO system in the  current time slot is given as
\begin{align}
\begin{aligned}
&\left[ {\begin{array}{*{20}{c}}
	{N_1^{R{x_1}}}\\
	{N_1^{R{x_2}}}
	\end{array}} \right]=\\
& \left[ {\begin{array}{*{20}{c}}
	{\delta _1^{(1,{x_j})}}&{\delta _1^{(2,{x_j})}}\\
	{\delta _1^{(2,{x_j})}}&{\delta _1^{(1,{x_j})}}
	\end{array}} \right]\left[ {\begin{array}{*{20}{c}}
	{{a_1}}&0\\
	0&{{a_2}}
	\end{array}} \right]\left[ {\begin{array}{*{20}{c}}
	{{x_1}}\\
	{{x_2}}
	\end{array}} \right] + \left[ {\begin{array}{*{20}{c}}
	{{I_1}}\\
	{{I_2}}
	\end{array}} \right],
\end{aligned}
\end{align}
or
\begin{align}
\begin{aligned}
\left[ {\begin{array}{*{20}{c}}
	{N_1^{R{x_1}}}\\
	{N_1^{R{x_2}}}
	\end{array}} \right] = \left[ {\begin{array}{*{20}{c}}
	{\delta _1^{(1,{x_j})}}&{\delta _1^{(2,{x_j})}}\\
	{\delta _1^{(2,{x_j})}}&{\delta _1^{(1,{x_j})}}
	\end{array}} \right]\left[ {\begin{array}{*{20}{c}}
	{{a_1}{x_1}}\\
	{{a_2}{x_2}}
	\end{array}} \right] + \left[ {\begin{array}{*{20}{c}}
	{{I_1}}\\
	{{I_2}}
	\end{array}} \right],
\label{ChannelModel}
\end{aligned}
\end{align}
where $x_j$ can get $L_0$ or $L_1$ according to the bit of information which is sent by the $j$th active transmitter in current time slot.  In the equations above,  the symmetric structure  for molecular communication system is considered, i.e.,
\begin{align}
\begin{aligned}
&\delta_1^{(1,{x_j})} \buildrel \Delta \over = \delta_1^{(2,2,{x_j})} = \delta_1^{(1,1,{x_j})},\\
&\delta_1^{(2,{x_j})} \buildrel \Delta \over = \delta_1^{(2,1,{x_j})} = \delta_1^{(1,2,{x_j})},\\
\end{aligned}
\end{align}
are used.
To make \eqref{ChannelModel}  simpler, we use
\begin{align}
\begin{aligned}
&\bm{y}= \bm{H}\bm{x} + \bm{i},
\end{aligned}
\end{align}
in which matrix $\bm{H}$  is a channel matrix and vectors  $\bm{x}$, $\bm{y}$ and $\bm{i}$ are the number of transmitted molecules by each of transmitters, the number of received molecules by each of receivers and the ISI vector, respectively.
\subsubsection{Detection}
In this section, the match filter (MF) detector for this 2$\times$2 MIMO system is proposed as follows
\begin{align}
\begin{aligned}
\bm{\hat y} = {{\bm{\bar H}}^{ - 1}}\bm{y},
\label{DETECT}
\end{aligned}
\end{align}
inspired by the detection algorithm in \cite{MIMO_prototype} and detection methods in radio-based communication.
According to \eqref{DETECT},  the average channel matrix in this method is needed.
It should be mentioned that the channel matrix in the proposed method completely  differs from the method in \cite{MIMO_prototype}.\\
Using \eqref{SigmaBino}, the average channel matrix in \eqref{ChannelModel} is calculated as
\begin{align}
\begin{aligned}
\bm{\bar H}_{2\times 2} = \left[ {\begin{array}{*{20}{c}}
	{p_1^{(1)}}&{p_1^{(2)}}\\
	{p_1^{(2)}}&{p_1^{(1)}}
	\end{array}} \right],
\label{meanH}
\end{aligned}
\end{align}
where ${p_1^{(i)}}$ is th mean value of random variable $\delta_1^{(i,{x_j})}$. According to \ref{III}  this matrix is full rank and invertible. Considering \eqref{DETECT} and utilizing \eqref{meanH} and \eqref{ChannelModel} provides the following
\begin{align}
\begin{aligned}
\bm{ \hat y} = K\left[\hspace{-1mm}{\begin{array}{*{20}{c}}
	{p_1^{(1)}}&{ - p_1^{(2)}}\\
	{ - p_1^{(2)}}&{p_1^{(1)}}
	\end{array}}\hspace{-1mm}\right]\hspace{-2mm}\left[ {\begin{array}{*{20}{c}}
	{{a_1}\delta _1^{(1,{x_1})}{x_1} + {a_2}\delta _1^{(2,{x_2})}{x_2} + {I_1}}\\
	{{a_1}\delta _1^{(2,{x_1})}{x_1} + {a_2}\delta _1^{(1,{x_2})}{x_2} + {I_2}}
	\end{array}} \right],
\end{aligned}
\end{align}
where
\begin{align}
\begin{aligned}
K \buildrel \Delta \over = \frac{1}{{{{(p_1^{(1)})}^2} - {{(p_1^{(2)})}^2}}}.
\label{K}
\end{aligned}
\end{align}
Because of the symmetric topology of this MIMO system it is enough to only analyze $Rx_1$.  So, the detector output at the $Rx_1$ side is
\begin{align}
\begin{aligned}
{\hat y(1)} = &K[{a_1}(p_1^{(1)}\delta _1^{(1,{x_1})} - p_1^{(2)}\delta _1^{(2,{x_1})}){x_1}\\
&+ {a_2}(p_1^{(1)}\delta _1^{(2,{x_2})} - p_1^{(2)}\delta _1^{(1,{x_2})}){x_2} + p_1^{(1)}{I_1} - p_1^{(2)}{I_2})].
\end{aligned}
\end{align}
According to this equation the detector output at the $Rx_1$ side can be obtained given the $j$th transmitter is active and sending bit 0 or bit 1. As a result the detector output has four states
\begin{align}
\begin{aligned}
&\hat y(1){|_{00}}\hspace{-1mm}=\hspace{-1mm}K[(p_1^{(1)}\hspace{-1mm}\delta _1^{(1,{L_0})}\hspace{-1mm}{L_0} - p_1^{(2)}\hspace{-1mm}\delta _1^{(2,{L_0})}\hspace{-1mm}{L_0}) + p_1^{(1)}\hspace{-1mm}{I_1} - p_1^{(2)}\hspace{-1mm}{I_2})],\\
&\hat y(1){|_{01}}\hspace{-1mm}=\hspace{-1mm}K[(p_1^{(1)}\hspace{-1mm}\delta _1^{(1,{L_1})}\hspace{-1mm}{L_1} - p_1^{(2)}\hspace{-1mm}\delta _1^{(2,{L_1})}\hspace{-1mm}{L_1}) + p_1^{(1)}\hspace{-1mm}{I_1} - p_1^{(2)}\hspace{-1mm}{I_2})],\\
&\hat y(1){|_{10}}\hspace{-1mm}=\hspace{-1mm}K[(p_1^{(1)}\hspace{-1mm}\delta _1^{(2,{L_0})}\hspace{-1mm}{L_0} - p_1^{(2)}\hspace{-1mm}\delta _1^{(1,{L_0})}\hspace{-1mm}{L_0}) + p_1^{(1)}\hspace{-1mm}{I_1} - p_1^{(2)}\hspace{-1mm}{I_2})],\\
&\hat y(1){|_{11}}\hspace{-1mm}=\hspace{-1mm}K[(p_1^{(1)}\hspace{-1mm}\delta _1^{(2,{L_1})}\hspace{-1mm}{L_1} - p_1^{(2)}\hspace{-1mm}\delta _1^{(1,{L_1})}\hspace{-1mm}{L_1}) + p_1^{(1)}\hspace{-1mm}{I_1} - p_1^{(2)}\hspace{-1mm}{I_2})].
\label{yhat}
\end{aligned}
\end{align}
In these equations $\delta _1^{(i,{L_1})}{L_1}$ and $\delta _1^{(i,{L_0})}{L_0}$ are Binomial random variables.
\\Generally if $U$ and $V$ follow
${\mathcal {N}}(\mu_U,\sigma^2_U)$ and
${\mathcal {N}}(\mu_V,\sigma^2_V)$ respectively, it can be proved \cite{popul},
\begin{align}
\begin{aligned}
(U + \alpha V) \sim \mathcal{N}({\mu _{_U}} + \alpha {\mu _{_V}},\sigma _{_U}^2 + {(\alpha {\sigma _{_V}})^2}),
\label{UV}
\end{aligned}
\end{align}
where $\alpha$ is a deterministic variable.
Using \eqref{UV} and \eqref{yhat} the mean of detector output which follows Normal distribution is
\begin{align}
\begin{aligned}
&{\mu _{\hat y(1){|_{00}}}} = K[({(p_1^{(1)})^2} - {(p_1^{(2)})^2}){L_0} +C_1 ],\\
&{\mu _{\hat y(1){|_{01}}}} = K[({(p_1^{(1)})^2} - {(p_1^{(2)})^2}){L_1} + C_1],\\
&{\mu _{\hat y(1){|_{10}}}} = K{C_1},\\
& {\mu _{\hat y(1){|_{11}}}} = K{C_1},
\end{aligned}
\end{align}
and its variance is
\begin{align}
\begin{aligned}
\sigma _{\hat y(1){|_{00}}}^2\hspace{-1mm}&=\hspace{-1mm} {K^2}[({(p_1^{(1)})^3}(1 - p_1^{(1)}) + {(p_1^{(2)})^3}(1 - p_1^{(2)})){L_0} + C_2],\\
\sigma _{\hat y(1){|_{01}}}^2\hspace{-1mm}&=\hspace{-1mm} {K^2}[({(p_1^{(1)})^3}(1 - p_1^{(1)}) + {(p_1^{(2)})^3}(1 - p_1^{(2)})){L_1} + C_2],\\
\sigma _{\hat y(1){|_{10}}}^2\hspace{-1mm}&=\hspace{-1mm} {K^2}[({(p_1^{(1)})^2}p_1^{(2)}(1 - p_1^{(2)}) + {(p_1^{(2)})^2}p_1^{(1)}(1 - p_1^{(1)})){L_0} \\
&+ C_2],\\
\sigma _{\hat y(1){|_{11}}}^2\hspace{-1mm}&=\hspace{-1mm} {K^2}[({(p_1^{(1)})^2}p_1^{(2)}(1 - p_1^{(2)}) + {(p_1^{(2)})^2}p_1^{(1)}(1 - p_1^{(1)})){L_1} \\
&+ C_2],
\end{aligned}
\end{align}
where
\begin{align}
\begin{aligned}
&C_1=(p_1^{(1)} - p_1^{(2)}){\mu_I},\\
&C_2=({(p_1^{(1)})^2} + {(p_1^{(2)})^2})\sigma _I^2,
\end{aligned}
\end{align}
and
\begin{align}
\begin{aligned}
{\mu _I} &\buildrel \Delta \over = {\mu _{{I_1}}} = {\mu _{{I_2}}} = \frac{1}{4}({L_0} + {L_1})(p_2^{(1)} + p_2^{(2)}),\\
\sigma _I^2  &\buildrel \Delta \over = \sigma _{{I_1}}^2 = \sigma _{{I_2}}^2\\&= \frac{1}{16}({L_0} + {L_1})(p_2^{(1)}(1 - p_2^{(1)}) + p_2^{(2)}(1 - p_2^{(2)})).
\end{aligned}
\end{align}
In summary, at the end of each time slot, we define the number of captured molecules by each receiver as $\bm{y}$ and $\bm{\hat y}$ is the product of the inverse channel matrix and $\bm{y}$.  According to section \ref{III}, we can use diversity scheme in this 2$\times$2 MIMO system. So, we define
\begin{align}
\begin{aligned}
\hat y_{Sub}\triangleq \hat y(1)-\hat y(2),
\end{aligned}
\end{align}
which is the difference between the detector outputs in two receivers.
Therefore, the mean and the variance of $\hat y_{Sub}$ for a given state of the transmitted symbol in each time slot can be calculated.
Using these values, three threshold levels are obtained. By finding the decision rule, the detection algorithm ends completely. Note that by defining $\hat y_{Sub}$ as above, the mean and the variance of ISI will be removed from the mean and the variance of $\hat y_{Sub}$ equations respectively and the detection process will go on easier. By any other linear combination of $\hat y(1)$ and $\hat y(2)$ this result will not be achieved.
\subsection{Analysis of  4$\times$4 MSM System}
\subsubsection{Proposed modulation scheme}
In this section, a 4$\times$4 MIMO system is introduced to increase the data rate. The more symmetric the molecular MIMO system is, the easier the analysis and the design of detector will be. Thus, we do not use any arbitrary topology, to increase the number of transmitters and receivers.  \figref{asd} shows the suggested 4$\times$4 molecular MIMO system. In this symmetric topology all of the receivers and transmitters are located on the vertices of one virtual cuboid and it is enough to only analyze $Rx_1$. Using MSM two bits are needed to enumerate all the transmitters. As a result, three bits are transmitted in each time slot. As we did before, the molecule hitting probability in this system is calculated as \eqref{controli}. The number of received molecules at the $Rx_i$ is given in \eqref{chMat4x4}.
All variables are defined the same as before.
\begin{figure*}[!t]
	\normalsize
	\begin{equation}
	\begin{aligned}
	\left[ {\begin{array}{*{20}{c}}
		{N_1^{R{x_1}}}\\
		{N_1^{R{x_2}}}\\
		{N_1^{R{x_3}}}\\
		{N_1^{R{x_4}}}
		\end{array}} \right] = \left[ {\begin{array}{*{20}{c}}
		{\delta _1^{(1,{x_j})}}&{\delta _1^{(2,{x_j})}}&{\delta _1^{(3,{x_j})}}&{\delta _1^{(4,{x_j})}}\\
		{\delta _1^{(2,{x_j})}}&{\delta _1^{(1,{x_j})}}&{\delta _1^{(4,{x_j})}}&{\delta _1^{(3,{x_j})}}\\
		{\delta _1^{(3,{x_j})}}&{\delta _1^{(4,{x_j})}}&{\delta _1^{(1,{x_j})}}&{\delta _1^{(2,{x_j})}}\\
		{\delta _1^{(4,{x_j})}}&{\delta _1^{(3,{x_j})}}&{\delta _1^{(2,{x_j})}}&{\delta _1^{(1,{x_j})}}
		\end{array}} \right]\left[ {\begin{array}{*{20}{c}}
		{{a_1}}&0&0&0\\
		0&{{a_2}}&0&0\\
		0&0&{{a_3}}&0\\
		0&0&0&{{a_4}}
		\end{array}} \right]\left[ {\begin{array}{*{20}{c}}
		{{x_1}}\\
		{{x_2}}\\
		{{x_3}}\\
		{{x_4}}
		\end{array}} \right] + \left[ {\begin{array}{*{20}{c}}
		{{I_1}}\\
		{{I_2}}\\
		{{I_3}}\\
		{{I_4}}
		\end{array}} \right],
	\centering
	\label{chMat4x4}
	\end{aligned}
	\end{equation}
	\hrulefill
	\vspace*{4pt}
\end{figure*}
\subsubsection{Detection}
In this system, the transmitter sends three bits in each time slot. Thus, if the previous detection method is used, seven threshold levels are needed. Therefore, a new detection method is proposed for this scheme. For simplicity, we perform this method in two steps. The first step is the transmitter recognition from which the molecules are originated and the next step is the transmitted bit detection. If the length of the time slot is large enough and a significant number of molecules received at $Rx_i$ corresponding to the $Tx_i$, in the first step the total number of molecules which captured by each receiver is computed in one time slot.
The receiver which has captured more molecules than the others is marked. Therefor, in the current time slot we assume that the corresponding transmitter to the marked receiver has sent molecules; This is reasonable because the number of received molecules is inversely proportional to the distance between the transmitter and the receiver. Therefore, the transmitter is determined in the current time slot. In the next step, a decision should be made about the sent bit. Considering section \ref{III} results, we can use diversity in this system and utilize the number of captured molecules in all receivers to make a decision. At the receivers side the MF detector is used. Thus, the average of the channel matrix is needed. From \eqref{chMat4x4} the formulation
\begin{align}
\begin{aligned}
\bm{\bar H}_{4\times 4} = \left[ {\begin{array}{*{20}{c}}
	{p_1^{(1)}}&{p_1^{(2)}}&{p_1^{(3)}}&{p_1^{(4)}}\\
	{p_1^{(2)}}&{p_1^{(1)}}&{p_1^{(4)}}&{p_1^{(3)}}\\
	{p_1^{(3)}}&{p_1^{(4)}}&{p_1^{(1)}}&{p_1^{(2)}}\\
	{p_1^{(4)}}&{p_1^{(3)}}&{p_1^{(2)}}&{p_1^{(1)}}\\
	\end{array}} \right],
\end{aligned}
\end{align}
is achieved. This matrix is symmetric so its inverse matrix is symmetric as follows
\begin{align}
\begin{aligned}
{{\bm{\bar H}}^{ - 1}}_{4\times 4} = \frac{1}{{\det (\bm{\bar H_{4\times 4}})}}\left[ {\begin{array}{*{20}{c}}
	{{c_{11}}}&{{c_{12}}}&{{c_{13}}}&{{c_{14}}}\\
	{{c_{12}}}&{{c_{11}}}&{{c_{14}}}&{{c_{13}}}\\
	{{c_{13}}}&{{c_{14}}}&{{c_{11}}}&{{c_{12}}}\\
	{{c_{14}}}&{{c_{13}}}&{{c_{12}}}&{{c_{11}}}\\
	\end{array}} \right],
\label{ExpMat4}
\end{aligned}
\end{align}
Using \eqref{DETECT} and \eqref{ExpMat4}, the detector output at $Rx_1$ side is
\begin{align}
\begin{aligned}
\hat y(1) &=\frac{1}{{\det (\bm{\bar H_{4\times 4}})}}\times\\
&\bigg[{a_1}({c_{11}}\delta _1^{(1,{x_1})} + {c_{12}}\delta _1^{(2,{x_1})} + {c_{13}}\delta _1^{(3,{x_1})} + {c_{14}}\delta _1^{(4,{x_1})}){x_1}\\
& + {a_2}({c_{11}}\delta _1^{(2,{x_2})} + {c_{12}}\delta _1^{(1,{x_2})} + {c_{13}}\delta _1^{(4,{x_2})} + {c_{14}}\delta _1^{(3,{x_2})}){x_2}  \\
& +{a_3}({c_{11}}\delta _1^{(3,{x_3})} + {c_{12}}\delta _1^{(4,{x_3})} + {c_{13}}\delta _1^{(1,{x_3})} + {c_{14}}\delta _1^{(2,{x_3})}){x_3} \\
& + {a_4}({c_{11}}\delta _1^{(4,{x_4})} + {c_{12}}\delta _1^{(3,{x_4})} + {c_{13}}\delta _1^{(2,{x_4})} + {c_{14}}\delta _1^{(1,{x_4})}){x_4} \\&+
{c_{11}}{I_1} + {c_{12}}{I_2} + {c_{13}}{I_3} + {c_{14}}{I_4}\bigg].
\end{aligned}
\end{align}
Similarly, the detector output can be calculated for other receivers. Therefore, the mean of the detector outputs for different symbols can be obtained. For example for symbol 000 the means of the detector outputs are as \eqref{eq61}.
\begin{figure*}[!t]
	\normalsize
	\begin{align}
	\begin{aligned}
	\begin{array}{l}
	{\mu _{\hat y(1){|_{000}}}} = \frac{1}{{\det (\bm{\bar H_{4\times 4}})}}(({c_{11}}p_1^{(1)} + {c_{12}}p_1^{(2)} + {c_{13}}p_1^{(3)} + {c_{14}}p_1^{(4)}){L_0} + ({c_{11}} + {c_{12}} + {c_{13}} + {c_{14}}){\mu _{I'}}),\\
	{\mu _{\hat y(2){|_{000}}}} = \frac{1}{{\det (\bm{\bar H_{4\times 4}})}}(({c_{12}}p_1^{(1)} + {c_{11}}p_1^{(2)} + {c_{14}}p_1^{(3)} + {c_{13}}p_1^{(4)}){L_0} + ({c_{11}} + {c_{12}} + {c_{13}} + {c_{14}}){\mu _{I'}}),\\
	{\mu _{\hat y(3){|_{000}}}} = \frac{1}{{\det (\bm{\bar H_{4\times 4}})}}(({c_{13}}p_1^{(1)} + {c_{14}}p_1^{(2)} + {c_{11}}p_1^{(3)} + {c_{12}}p_1^{(4)}){L_0} + ({c_{11}} + {c_{12}} + {c_{13}} + {c_{14}}){\mu _{I'}}),\\
	{\mu _{\hat y(4){|_{000}}}} = \frac{1}{{\det (\bm{\bar H_{4\times 4}})}}(({c_{14}}p_1^{(1)} + {c_{13}}p_1^{(2)} + {c_{12}}p_1^{(3)} + {c_{11}}p_1^{(4)}){L_0} + ({c_{11}} + {c_{12}} + {c_{13}} + {c_{14}}){\mu _{I'}}).
	\end{array}
	\label{eq61}
	\end{aligned}
	\end{align}
	\hrulefill
	\vspace*{4pt}
\end{figure*}
Similarly, the variance of the detector outputs can be obtained as \eqref{eq63}.
\begin{figure*}[!t]
	\normalsize
\begin{align}
\begin{aligned}
\begin{array}{l}
\sigma _{\hat y(1){|_{000}}}^2 = \frac{1}{{{{(\det (\bm{\bar H_{4\times 4}}))}^2}}}(({c_{11}}^2p_1^{(1)}(1 - p_1^{(1)}) + {c_{12}}^2p_1^{(2)}(1 - p_1^{(2)}) +{c_{13}}^2p_1^{(3)}(1 - p_1^{(3)}) \\\hspace{1.8cm} +{c_{14}}^2p_1^{(4)}(1 - p_1^{(4)})){L_0} + ({c_{11}}^2 + {c_{12}}^2 + {c_{13}}^2 + {c_{14}}^2)\sigma _{I'}^2),\\\\
\sigma _{\hat y(2){|_{000}}}^2 = \frac{1}{{{{(\det (\bm{\bar H_{4\times 4}}))}^2}}}(({c_{12}}^2p_1^{(1)}(1 - p_1^{(1)}) + {c_{11}}^2p_1^{(2)}(1 - p_1^{(2)}) +{c_{14}}^2p_1^{(3)}(1 - p_1^{(3)})\\\hspace{1.8cm} + {c_{13}}^2p_1^{(4)}(1 - p_1^{(4)})){L_0} + ({c_{11}}^2 + {c_{12}}^2 + {c_{13}}^2 + {c_{14}}^2)\sigma _{I'}^2),\\\\
\sigma _{\hat y(3){|_{000}}}^2 = \frac{1}{{{{(\det (\bm{\bar H_{4\times 4}}))}^2}}}(({c_{13}}^2p_1^{(1)}(1 - p_1^{(1)}) + {c_{14}}^2p_1^{(2)}(1 - p_1^{(2)}) +{c_{11}}^2p_1^{(3)}(1 - p_1^{(3)})\\\hspace{1.8cm}  + {c_{12}}^2p_1^{(4)}(1 - p_1^{(4)})){L_0} +({c_{11}}^2 + {c_{12}}^2 + {c_{13}}^2 + {c_{14}}^2)\sigma _{I'}^2),\\\\
\sigma _{\hat y(4){|_{000}}}^2 = \frac{1}{{{{(\det (\bm{\bar H_{4\times 4}}))}^2}}}(({c_{14}}^2p_1^{(1)}(1 - p_1^{(1)}) + {c_{13}}^2p_1^{(2)}(1 - p_1^{(2)})  +{c_{12}}^2p_1^{(3)}(1 - p_1^{(3)})\\\hspace{1.8cm} + {c_{11}}^2p_1^{(4)}(1 - p_1^{(4)})){L_0} + ({c_{11}}^2 + {c_{12}}^2 + {c_{13}}^2 + {c_{14}}^2)\sigma _{I'}^2),
\end{array}
\label{eq63}
\end{aligned}
\end{align}
	\hrulefill
	\vspace*{4pt}
\end{figure*}
where
\begin{align}
\begin{aligned}
&{\mu _{I'}} = \frac{1}{8}({L_0} + {L_1})(p_2^{(1)} + p_2^{(2)} + p_2^{(3)} + p_2^{(4)}),\\
&\sigma _{I'}^2 = \frac{1}{{64}}({L_0} + {L_1})(p_2^{(1)}(1 - p_2^{(1)}) + p_2^{(2)}(1 - p_2^{(2)})\\
& + p_2^{(3)}(1 - p_2^{(3)}) + p_2^{(4)}(1 - p_2^{(4)})),
\end{aligned}
\end{align}
is the mean and the variance of ISI which is the same for all receivers.
Using diversity in this system we define
\begin{align}
\begin{aligned}
{\hat y}_{_{Sum}} = \sum\limits_{n = 1}^4 {\hat y(n)},
\label{Sum_Y}
\end{aligned}
\end{align}
which is the sum of the MF output of all receivers in the current time slot. According to the system, symmetry sending bit 0 or bit 1 is the only factor that changes the mean value of ${\hat y}_{_{Sum}}$. In other words,  ${\mu _{{{\hat y}_{sum}}}}$ will be the same for sending the symbols 001, 011, 101 and 111. Therefore, using \eqref{eq61}, \eqref{eq63} and \eqref{Sum_Y} two different values for the mean and the variance of ${\hat y}_{_{Sum}}$  are obtained and the intersection point of the two Normal distributions is calculated as the threshold level. Note that we used all the outputs of the receivers in the 4$\times$4 system. The reason is that there is not any special linear combination  that can remove ISI effect in the decision, like in 2$\times$2 systems; and, in comparison to other combinations, the used linear combination has a better result in simulation.
\section{Simulation results}
\begin{figure}
	\centering
	\includegraphics[width=1\columnwidth]{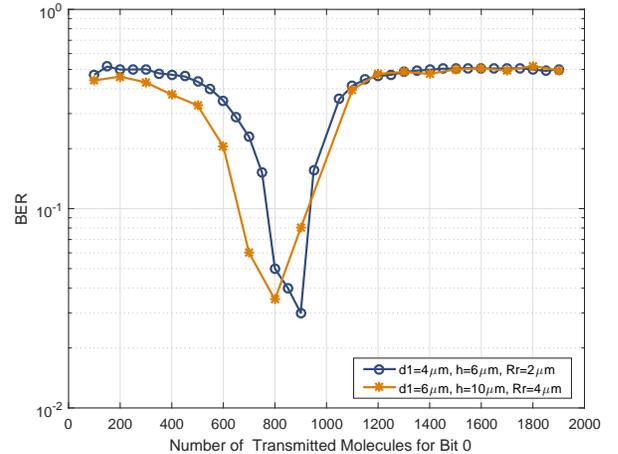}
	\caption{BER of 2$\times$1 system in terms of $L_0$ for two different configurations.}
	\label{BERvsQ0_final}
	\centering
\end{figure}

\begin{figure}
	\centering
	\includegraphics[width=1\columnwidth]{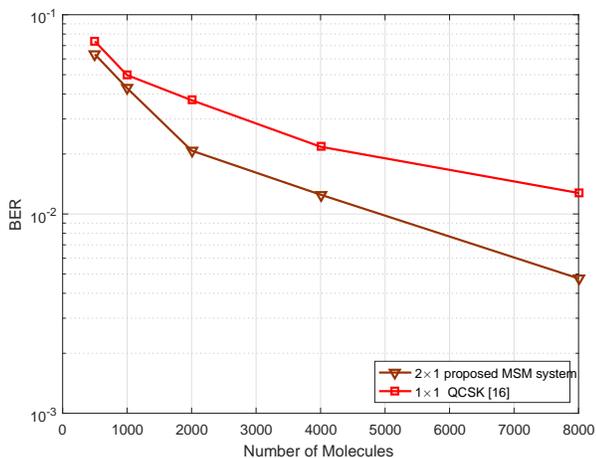}
	\caption{BER of 2$\times$1 system based on MSM and 1$\times$1 system based on QCSK in terms of average number of molecules.}
	\label{2TxVsQCSK}
	\centering
\end{figure}

\begin{figure}
	\centering
	\includegraphics[width=1\columnwidth]{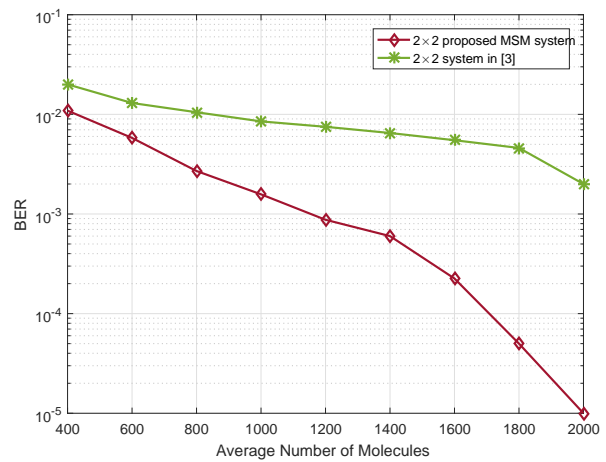}
	\caption{BER of 2$\times$2 system based on MSM and 2$\times$2 system based on proposed method in \cite{MIMO_prototype} in terms of average number of molecules.}
	\label{2Tx_2Rx_myMethod_Yilmaz_new2}
	\centering
\end{figure}

\begin{figure}
	\centering
	\includegraphics[width=1\columnwidth]{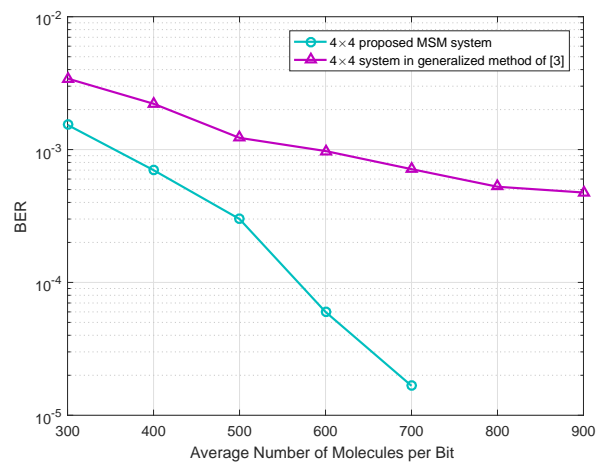}
	\caption{BER comparison in 4$\times$4 system between MSM and generalized method which is proposed in \cite{MIMO_prototype} in terms of average number of molecules per bit.}
	\label{BER4x4VsYilmaz}
	\centering
\end{figure}

We evaluate the performance of the proposed schemes. We compare BER of MSM systems to the other systems with similar data rates. We use MATLAB for these simulations and the algorithm \ref{alg:RANSAC} describes
the simulation pseudo-code which obtains the molecule absorption probability in the receivers. Also, the key parameters are summarized in table \ref{jad_param} which are constant for all simulations \cite{Andrews2012}, \cite{MIMO_prototype}.
\begin{table}[htp]
	\centering \caption{\footnotesize Key parameters in our simulation}
	\renewcommand*{\arraystretch}{2}
	\begin{tabular}{| c | c | }
		\hline
		Explenation & 	value \\
		\hline
		Diffusion coefficient ($D$)
		
		&
		50	${\raise0.7ex\hbox{${\mu {m^2}}$} \!\mathord{\left/
				{\vphantom {{\mu {m^2}} s}}\right.\kern-\nulldelimiterspace}
			\!\lower0.7ex\hbox{$s$}}$
		\\
		\hline
		Time step ($\Delta t$)
		&
		0.1$ms$
		\\
		\hline
		Time slot length ($T_s$)
		&
		$\{0.1 \sim 1\}s$
		\\
		\hline
	\end{tabular}
	\label{jad_param}
\end{table}
First, we simulate  2$\times$1 systems. In \figref{BERvsQ0_final}, the bit error rate of the 2$\times$1 system is shown in terms of the number of molecules emitted to send bit 0, i.e. $L_0$. In this figure, the average number of molecules sent over one time slot is 1000 molecules. It is seen that the number of  molecules to send bit 0 and bit 1 greatly affects BER, so the optimization problem introduced in \eqref{optimization} should be solved and the proper values for $L_0$ and $L_1$ be calculated.\\
In \cite{RN127} CSK modulation is introduced with four levels, also known as QCSK. Using this modulation, in the 1$\times$1 system two bits can be sent to the receiver at any time slot. In other words, for each of the symbols 00, 01, 10 and 11 a level is considered, and the different number of molecules for each of these symbols are used. In fact, this modulation is a generalized CSK modulation which is introduced to increase the data rate in  systems with one transmitter and receiver. Therefore, this method can be used as a method for comparing MSM modulation in a the 2$\times$1 system because they have the same data rates.\\
In \figref{2TxVsQCSK}, the BER of the 2$\times$1 system based on MSM  and 1$\times$1  system based on QCSK are shown in terms of the  average number of molecules emitted to send bit 0 and 1. It is shown that the MS modulation has a lower bit error rate than QCSK modulation.  The receiver radius and the transmitter distance from the receiver are equal to $d_1=4\mu m$ and $R_r=2\mu m$, respectively. The distance between $Tx_2$ and the receiver is computed  using  ${d_2} =\sqrt{h^{2}+{(d_1+R_r)}^{2}}-R_r$  and in MSM system  $Tx_2$ is located at 4$\mu m$ distance from the  $Tx_1$ based on \figref{2Tx}. Each time slot with a duration of about $100$\textit{ms} is considered and for ensuring  fair comparison, the average number of molecules  sent in two systems are considered equivalent.\\
For each average number of molecules, the number of molecules which are sent for bit 0 and 1 is calculated from convex optimization problem in \eqref{optimization}. The four quantity levels which are used for QCSK are uniformly spaced like \cite{On} and \cite{signal}. Note that in two systems the threshold level detection is used. \\
To show the performance of the MS modulation in the 2$\times$2 system we compare the BER of the system with the proposed scheme in \cite{MIMO_prototype}. The results are shown in \figref{2Tx_2Rx_myMethod_Yilmaz_new2}.
In \cite{MIMO_prototype}, in each time slot, each transmitter sends a bit to the receiver which is its pair source of communication. Therefore, the ILI occurs. Also, in this paper, the transmitter sends no molecules for sending bit 0.  To have fair comparison the average number of molecules which are sent in two systems is considered equivalent. To achieve the BER curve in \figref{2Tx_2Rx_myMethod_Yilmaz_new2} the distance between $Tx_1$ from $Rx_1$ and ${Tx}_2$ from $Rx_2$ is 6$\mu m$.  The two receivers have the same radius 2$\mu m$ , and are
placed $5\mu m$ distance apart and the time slot duration is about $1s$. It can be observed in  \figref{2Tx_2Rx_myMethod_Yilmaz_new2} that the BER of 2$\times$2 MSM system  is much lower than the system  proposed in \cite{MIMO_prototype}. For example, there is about one decade gap in the average number of $1400$ molecules. \\
We have generalized the proposed method in \cite{MIMO_prototype} for the 4$\times$4 molecular communication system. Then, we have compared BER performance of our MSM method with this work in \figref{BER4x4VsYilmaz}. In this system (\figref{asd}) the values of $d_1$, $w$, $h$, $R_r$ and $T_s$ are $6\mu m$, $4\mu m$, $10\mu m$, $3\mu m$ and $1s$ respectively.
\begin{figure}
	\centering
	\includegraphics[width=1\columnwidth]{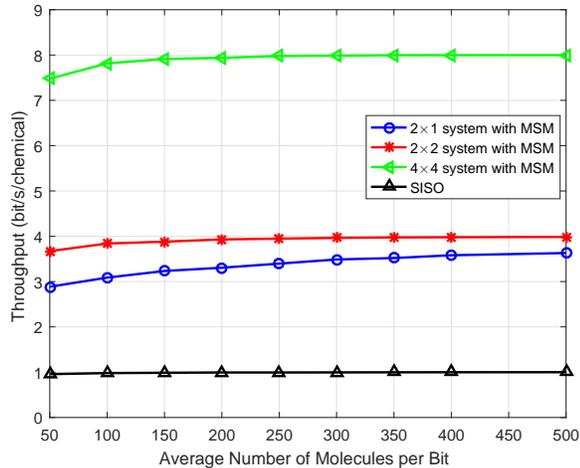}
	\caption{Throughput comparison in 4$\times$4, 2$\times$2, 2$\times$1 MSM systems and the SISO system ($T_S$=1s). }
	\label{Throughput}
	\centering
\end{figure}
As mentioned before in the 4$\times$4 MSM system, three bits are transmitted however in the generalized method of \cite{MIMO_prototype}, four bits are transmitted in each time slot. Therefore, for a fair comparison between two methods, we fix the average number of molecules per bit. As it can be seen in \figref{BER4x4VsYilmaz} the BER of MSM is lower than the generalized method of \cite{MIMO_prototype}. \\ As the last simulation, in \figref{Throughput} the throughput of 4$\times$4, 2$\times$2, 2$\times$1 MSM systems and the SISO system are compared. The throughput is calculated using $ \frac{{MN}}{{{T_s}}}(1 - BER)$ \cite{MIMO_prototype}, where $M$ and $N$ are the modulation order and the spatial streams respectively, where a spatial stream is the number of bit streams that can be sent simultaneously through a MIMO system. As expected the throughput of the 4$\times$4 system is higher than the other systems.
\section{Conclusion}
In this paper, molecular MIMO communication systems have been considered to increase the data rate. Firstly, a new model has been introduced for the channel matrix in 2$\times$2 and 4$\times$4 systems. Using simulation, the matrix rank in these two systems has been calculated for different dimensions. It has been observed that this matrix is always full rank given that for bit 0 and bit 1, a number of molecules are transmitted by the transmitter. Therefore, the diversity techniques can be used in these MIMO systems.
Then, a new modulation in molecular communication has been presented called molecular spatial modulation or MSM. In this modulation, only one transmitter starts sending a bit in each time slot and depending on which transmitter is transmitting, the different bit is detected in the receiver. Moreover, special detectors have been presented in each 2$\times$1, 2$\times$2 and 4$\times$4 systems. Those detectors are all based on threshold level detection. Because of the full rank channel matrix, we have also used diversity in 2$\times$2 and 4$\times$4 systems. Finally, the suggested modulation has been compared with the other presented methods in prior papers based on bit error rate. To ensure the fairness of the comparison, the average number of molecules which are sent in two systems is considered equivalent.\\
According to all of the results and the plots presented, bit error rate in the suggested method in this paper is less than prior methods. Besides, the suggested modulation is very simple which is important because, in molecular communication, transmitter and receiver are not able to do complicated tasks.


\begin{thebibliography}{1}
	
	\bibitem{RN109}
	T.~Nakano, A.~W. Eckford, and T.~Haraguchi, {\em Molecular communication}.
	\newblock Cambridge University Press, 2013.
	
	\bibitem{RN113}
	T.~Nakano, M.~J.~Moore, F.~Wei, A.~V.~Vasilakos, and J.~Shuai, ``Molecular
	communication and networking: opportunities and challenges,''{\em IEEE Transactions on NanoBioscience}, vol.~11, no.~2, pp.~135--148, 2012.
	
		
		\bibitem{RN111}
		M.~Pierobon, and I.~F.~Akyildiz, ``A physical end-to-end model for molecular communication in nanonetworks,'' {\em IEEE Journal on Selected Areas in Communications}, vol.~28, no.~4, pp.~602--611, 2010.
		
		\bibitem{RN112}
		A.~W.~Eckford, ``Nanoscale communication with brownian motion,'' in {\em Proceedings of 41st Annual Conference on Information Sciences and Systems (CISS)}, 2007, pp.~160--165.
	
		\bibitem{MIMO_prototype}
		B.~H. Koo, C.~Lee, H.~B. Yilmaz, N.~Farsad, A.~Eckford, and C.~B. Chae,
		``Molecular MIMO: From theory to prototype,'' {\em IEEE Journal on Selected
			Areas in Communications}, vol.~34, no.~3, pp.~600--614, 2016.		
	
	\bibitem{32}
	L.~S. Meng, P.~C. Yeh, K.~C. Chen, and I.~F. Akyildiz, ``MIMO communications
	based on molecular diffusion,'' in {\em Proceedings of IEEE Global
		Telecommunications Conference (GLOBECOM)}, 2012, pp.~5380--5385.

	\bibitem{effect}
	M.~k. Kuran, H.~B. Yilmaz, T.~Tugcu, and I.~F. Akyildiz, ``Interference effects
	on modulation techniques in diffusion based nanonetworks,'' {\em Nano
		Communication Networks}, vol.~3, no.~1, pp.~65--73, 2012.
	\bibitem{2recv}
	Y.~Lu, M.~D. Higgins, A.~Noel, M.~S. Leeson, and Y.~Chen, ``The effect of two
	receivers on broadcast molecular communication systems,'' {\em IEEE
		Transactions on NanoBioscience}, vol.~15, no.~8, pp.~891--900, 2016.
	\bibitem{monte}
	D.~Arifler and D.~Arifler, ``Monte carlo analysis of molecule absorption
	probabilities in diffusion-based nanoscale communication systems with
	multiple receivers,'' {\em IEEE Transactions on NanoBioscience}, vol.~16,
	no.~3, pp.~157--165, 2017.
	
	\bibitem{coop}
	Y.~Fang, A.~Noel, N.~Yang, A.~W.~Eckford and R.~A.~Kennedy, ``Distributed cooperative detection for multi-receiver molecular communication,'' in {\em Proceedings of IEEE Global Telecommunications Conference (GLOBECOM)}, 2016.
	
	\bibitem{optimize}
	Y.~Fang, A.~Noel, N.~Yang, A.~W.~Eckford and R.~A.~Kennedy, ``Convex optimization of distributed cooperative detection in multi-receiver molecular communication,'' {\em IEEE Transactions on Molecular, Biological and Multi-Scale Communications}, 2018.
	
	\bibitem{learning}
	C.~Lee, H.~B.~Yilmaz, C.-B.~Chae, N.~Farsad, and A.~Goldsmith,
	``Machine learning based channel modeling for molecular MIMO communications,'' in {\em  Proceedings of IEEE 18th International Workshop on Signal
		Processing Advances in Wireless Communication (SPAWC)}, 2017, pp.~1--5.
	
	
	\bibitem{spatial}
	M.~Damrath, H.~B.~Yilmaz, C.-B.~Chae, and P.~A.~Hoeher, ``Spatial
	coding techniques for molecular MIMO,'' in {\em Proceedings of IEEE Information
		Theory Workshop (ITW)}, 2017, pp.~324--328.
	
	\bibitem{synch}
	Z.~Luo, L.~Lin, W.~Guo, S.~Wang, F.~Liu, and H~Yan, ``One symbol blind synchronization in SIMO molecular communication systems,'' {\em IEEE Wireless Communication Letter}, no.~99, pp.~1--1, 2018.
	
	
	\bibitem{CIRestim}
	S.~M.~Rouzegar and U.~Spagnolini, ``Channel estimation for diffusive MIMO molecular communications,'' in {\em Proceedings of European Conference on Networks
	and Communications (EuCNC)},  2017, pp.~1--5.
	
	\bibitem{SMMC}
	Y.~Huang, M.~Wen, L.~Yang, C.~B.~Chae and F.~Ji. ``Spatial Modulation for Molecular Communication.'', {\em arXiv preprint arXiv:1807.01468}, 2018.
	
	\bibitem{Communication over Diffusion-Based}
	Y.~Murin, N.~Farsad, M.~Chowdhury and A.~Goldsmith, ``Communication over diffusion-based molecular timing channels.'', in {\em Proceedings of Global Communications Conference (GLOBECOM)}, 2016.
	
	\bibitem{comprehen}
	N.~Farsad, H.~B. Yilmaz, A.~Eckford, C.~B. Chae, and W.~Guo, ``A comprehensive
	survey of recent advancements in molecular communication,'' {\em IEEE
		Communications Surveys \& Tutorials}, vol.~18, no.~3, pp.~1887--1919, 2016.
	
	\bibitem{3dim}
	H.~B.~Yilmaz, A.~C.~Heren, T.~Tugcu, and C.~B.~Chae, ``Three-dimensional channel characteristics for molecular communications with an absorbing receiver,'' {\em IEEE Communication Letters}, vol.~18, no.~6, pp.~929--932, Jun.~2014.
	
	
	
	\bibitem{Papoulis2002}
	A.~Papoulis and S.~U. Pillai, {\em Probability, random variables, and
		stochastic processes}.
	\newblock Tata McGraw-Hill Education, 2002.
	
	
	\bibitem{SpatialRef}
	R.~Y.~Mesleh, H.~Haas, S.~Sinanovic, C.~W.~Ahn and S.~Yun, ``Spatial modulation,'' {\em IEEE Transactions on Vehicular Technology}, vol.~57, no.~4, pp.~2228--2241, 2008.
	

	\bibitem{energy}
	M.~k. Kuran, H.~B. Yilmaz, T.~Tugcu, and B.~Ozerman, ``Energy model for
	communication via diffusion in nanonetworks,'' {\em Nano Communication
		Networks}, vol.~1, no.~2, pp.~86--95, 2010.
	
	\bibitem{RN114}
	N.~R.~Kim, A.~W.~Eckford, and C.~B.~Chae, ``Symbol interval optimization for molecular communication with drift,'' {\em IEEE Transaction on Nanobioscience}, vol.~13, no.~3, pp.~223--229, 2014.
	
	
	\bibitem{MoMotion}
	D.~T.~Gillespie and E.~Seitaridou, {\em Simple Brownian Diffusion: An Introduction to the Standard
	Theoretical Models}, Oxford University Press, 2012.
	
	
	\bibitem{Arj}
	H.~Arjmandi, A.~Gohari, M.~N. Kenari, and F.~Bateni, ``Diffusion-based
	nanonetworking: A new modulation technique and performance analysis,'' {\em
		IEEE Communications Letters}, vol.~17, no.~4, pp.~645--648, 2013.
	\bibitem{signal}
	W.~A. Lin, Y.~C. Lee, P.~C. Yeh, and C.~h. Lee, ``Signal detection and ISI
	cancellation for quantity-based amplitude modulation in diffusion-based
	molecular communications,'' in {\em Proceedings of Global Communications Conference
		(GLOBECOM)}, 2012, pp.~4362--4367.
	
	\bibitem{popul}
	A.~Papoulis, {\em Probability \& statistics}. \newblock Prentice-Hall Englewood Cliffs, 1990.
	
	\bibitem{Andrews2012}
	S.~S. Andrews, ``Spatial and Stochastic Cellular Modeling with the Smoldyn
	Simulator" in {\em Bacterial Molecular Networks: Methods and Protocols}, Springer New York, 2012, pp.~519--542.
	
	%
	%
	
	\bibitem{RN127}
	M.~S. Kuran, H.~B. Yilmaz, T.~Tugcu, and I.~F. Akyildiz, ``Modulation
	techniques for communication via diffusion in nanonetworks,'' in {\em Proceedings of
		IEEE International Conference on Communications (ICC)}, 2011, pp.~1--5.
	
	\bibitem{On}
	M.~U. Mahfuz, D.~Makrakis, and H.~T. Mouftah, ``On the characteristics of
	concentration-encoded multi-level amplitude modulated unicast molecular
	communication,'' in {\em Proceedings of 24th Canadian Conference on Electrical and
		Computer Engineering (CCECE)}, 2011, pp.~312--316.
	
\end{thebibliography}
\end{document}